\documentstyle[aps,prb,epsf,twocolumn]{revtex}

\begin{document}
\draft \twocolumn[\hsize\textwidth\columnwidth\hsize\csname
@twocolumnfalse\endcsname

\title{A Two-dimensional Infinte System Density Matrix Renormalization
Group Algorithm} \author{Patrik Henelius\cite{pa}} \address{National High
Magnetic Field Laboratory, Tallahassee, FL 32310 \\
Department of Physics, Indiana University, Bloomington, Indiana 47405}

\date{\today}  
\maketitle

\begin{abstract}

It has proved difficult to extend the density matrix renormalization
group technique to large two-dimensional systems. In this
Communication I present a novel approach where the calculation is done
directly in two dimensions. This makes it possible to use an infinite
system method, and for the first time the fixed point in two
dimensions is studied. By analyzing several related blocking schemes I
find that there exists an algorithm for which the local energy
decreases monotonically as the system size increases, thereby showing
the potential feasibility of this method.

\end{abstract}
\vskip2mm]

There is a great need for improved numerical techniques that are able
to treat two-dimensional electronic systems. Fundamental questions
such as the existence of superconductivity in the 2D Hubbard and t-J
models have not been resolved with present-day analytical or numerical
techniques.  All the commonly used numerical techniques suffer from
shortcomings: exact diagonalization is limited to small lattice sizes
due to the exponential growth of states with the system
size.\cite{Dago} Quantum Monte Carlo calculations are unable to reach
large fermion systems at low temperatures due to the ``sign
problem''.\cite{Loh} DMRG calculations have been very successful at
treating one-dimensional systems,\cite{White1} but accurate results
are difficult to obtain for large two-dimensional systems.

It seems most likely that major improvements towards a reliable 2D
technique will be made within the DMRG context. Years of effort have
resulted in no progress towards solving the fermion ``sign
problem''.\cite{Wies} It also seems unlikely that the size of
available computer memory will increase fast enough to facilitate
exact diagonalization calculations for large systems. Recent DMRG
studies,\cite{White3,Jongh} on the other hand, have reached the
largest 2D systems reported to date. In this Communication I approach
the 2D DMRG calculation from a new angle, which hopefully may
encourage further research in this direction.

First the basis of standard 1D DMRG and previous 2D DMRG calculations
will be reviewed. The source of the difficulties with previous 2D
calculations is discussed. Thereafter several new 2D blocking schemes
are proposed and tested, keeping only a small number of states per
block. Finally a promising algorithm is discussed in more detail. All
calculations are performed on the 2D antiferromagnetic Heisenberg
model. The ground state parameters of this fundamental
quantum-mechanical model are known to a high accuracy,\cite{Sand}
which makes it an ideal testing ground for new numerical techniques.

The central idea in a DMRG calculation is to iteratively increase the
system size, but to avoid an exponentially increasing number of states
by keeping only a fixed number of the ``most important'' states at
each iteration. In early numerical renormalization calculations the
lowest eigenstates of the system were chosen to be the ``important
states'', but the results were discouraging.\cite{White4} The major
breakthrough came with White's insight~\cite{White1} to use the
density matrix to determine which states to keep. In the superblock
method a number of blocks are combined together to form a
superblock. The superblock is divided into two parts, the ``system
block'' and the ``environment block''. At each iteration the
superblock is diagonalized and the density matrix is formed for the
system block. The density matrix is diagonalized and the importance of
each eigenstate is given by its eigenvalue. The states with the
largest eigenvalues are kept and the rest discarded. In the next step
of the iteration this system block, with a reduced number of states,
will be used in forming the new superblock. Thus the number of sites
increases with each iteration, while the number of states kept remains
constant.

\begin{figure}
\centering
\epsfxsize=8cm
\leavevmode
\epsffile{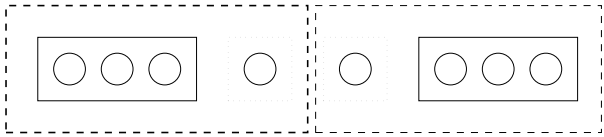}
\vskip0.5cm
\caption{Standard 1D blocking algorithm. The superblock consists of
the system block (heavy dashed line) and the environment block (dashed
line). The building block is indicated by a dotted line.}
\label{fig:1D}
\end{figure}

Using this basic formula many different DMRG algorithms can be
defined. An algorithm is characterized by how the superblock is
constructed, and by the manner in which the blocks are enlarged. The
most commonly used method was proposed in White's original work. The
superblock consists of four blocks, with the two central blocks
consisting of single sites, and the two end blocks being reflections
of each other, see Fig.~\ref{fig:1D}. The system block is taken to be
half the superblock, that is, one end block and an adjacent single
site, here called a building block. Using the density matrix a fixed
number of states are kept for the system block, which in the next
iteration will be recombined with a building block to form a new
system block. In this manner the system blocks are enlarged by the
size of the building block (here one site) at each iteration.  A
variation of this method is to form the superblock out of three
blocks, two end blocks and a single site in the middle. The system
block is chosen as above, but the environment block consists of a
single end block.

If the above procedure is iterated repeatedly one can reach
arbitrarily large systems. This is called the infinite system
method. In the thermodynamic limit the energy does approach a fixed
point, and as the number of states kept is increased, the fixed point
approaches the bulk ground state energy for the model. Usually the
infinite system method has been used to measure various quantities in
the middle of very large systems. In this manner results with an
accuracy of up to 13 digits have been reported.\cite{White2}

The rate of convergence does depend on how the superblock is
constructed. If one tries to add the building block at the outer
boundary of the superblock, then the local energy will increase as one
increases the system size, indicating that a good basis has not been
chosen. This can easily be understood since with open periodic
boundary conditions the wave function has to vanish at the
boundary. If one adds a site to the boundary it is clear that some
artifact will remain in the wave function as the system size is
increased. This somewhat trivial example shows that one cannot
construct an arbitrary blocking scheme and expect that a DMRG
calculation will yield a fast convergence. This will become more
evident for 2D systems.

In addition to the infinite system method discussed above, a finite
system method, also introduced in White's pioneering work, is commonly
used.  This algorithm is used to calculate properties of finite size
systems to high accuracy. Initially the infinite system method is used
to reach a system of desired length $L$. At each iteration the system
block is saved, so that when a system of size $L$ is reached, system
blocks of sizes 1 to $L/2-1$ are saved. Once the desired total system
size has been reached the superblock size is fixed. Next the saved
system blocks are used as environmental blocks while the system block
size is increased until it has reached the maximum length $L-3$. Now
system blocks of sizes 1 to $L-3$ are saved and these blocks can be
used as environment blocks for a consequent sweep through the lattice.
In this manner the basis kept in the system blocks can be iteratively
improved until convergence is reached.

\begin{figure}
\centering
\epsfxsize=3cm
\leavevmode
\epsffile{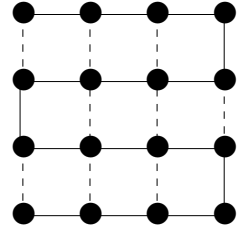}
\vskip0.5cm
\caption{Mapping of the 2D lattice to a 1D lattice with long-range 
interactions, the solid line indicates the 1d system, while
the dashed lines show the long-range interactions.}
\label{fig:2D}
\end{figure}

After this brief review of 1D DMRG calculations previous 2D
calculations will be considered next. Most previous 2D calculations
involve mapping the 2D lattice onto a 1D system with long-range
interactions,\cite{White3,Liang,Xiang,White5} see
Fig.~\ref{fig:2D}. Thereafter the above 1D finite system algorithm is
used. Notice that using this mapping it is not possible to use an
infinite system algorithm, since one determines the size of the final
lattice when doing the mapping. Also, the blocks break the symmetry of
the lattice and it is generally not possible to use a reflection of
the system block as the environment block. Therefore one has to use
some different trick to form the environment block for the initial
sweep. The two simplest options are to either leave the environment
block empty, or set all long-range interactions to zero in the initial
sweep. This method has, however, been able to treat the largest 2D
fermion systems to date, up to sizes 11 by 16.\cite{White3} An
alternative approach is to add a row of sites at each iteration.  In
this manner strips with a width of up to 6 sites and a length of 30
sites have been studied.\cite{Jongh}

Why have larger systems not been studied ? Liang and Pang~\cite{Liang}
found that for a 2D gas of free electrons, the number of states needed
to maintain a certain accuracy grows exponentially with the linear
system size. This convergence was also confirmed for an algorithm were
a row of sites was added at each step.\cite{Jongh} Although no proof
has been given, this statement is often referred to as most probably
valid for any 2D DMRG calculation. This statement was, however, made
for small finite size systems and it is not clear that it will apply
to possible infinite system methods. In an infinite system DMRG
calculation a fixed number of states are kept as the size of the
system is increased.  According to the above statement accuracy should
be lost in the process. For a system with open boundary conditions the
local energy decreases as the system size is increased. Furthermore,
due to the variational character of the technique,\cite{Ostl} the DMRG
energy is an upper bound on the energy of the system. Therefore
accuracy would certainly be lost if the DMRG energy increased as the
system size is increased, in agreement with the above statement. But
if the energy decreased as the system size is increased, then the
bound on the system energy is continuously improved, and in the limit
of the fixed point the relative accuracy will approach a constant
although only a fixed number of states is kept.

It was therefore the goal of this study to investigate whether there
exist 2D blocking algorithms for which the energy decreases
monotonically as the system size is increased. The algorithm should
retain more of the symmetry of the lattice so that it can be used in
the infinite system mode, and the fixed point studied directly. The
reason for the above mapping of the 2D system to a 1D system is that
it is not trivial to construct such an algorithm. Since the superblock
of most symmetric 2D algorithms is bound to consist of more blocks
than in the 1D case computer memory limitations will also be more
severe.

In order to build up a two-dimensional lattice in a more symmetric
fashion it seems likely that one has to use building blocks consisting
of several sites.  The simplest idea is probably to use a row of sites
as building block at each iteration.\cite{Jongh} This method has,
however, two shortcomings; it only grows the lattice in one spatial
dimension, and the number of states in the added row increases
exponentially with the width of the lattice.

\begin{figure}
\centering
\epsfxsize=6cm
\leavevmode
\epsffile{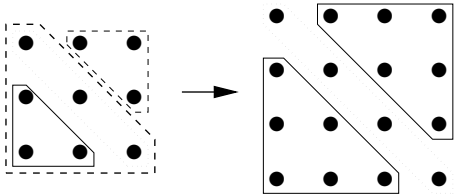}
\vskip0.5cm
\caption{A blocking algorithm for the 2D lattice where the system block
(heavy dashed line) is formed by combining the lower triangular part 
of the lattice with the diagonal block ( dotted line). The upper
triangular part of the lattice is used as environment block (light
dashed line).}
\label{fig:2D.1}
\end{figure}

In a first attempt to overcome these problems I divided the square 2D
lattice up into three blocks, consisting of the diagonal, a triangular
block below the diagonal, and the reflection of this block above the
diagonal, see Fig.~\ref{fig:2D.1}. The lower triangular block is used
as the system block, the diagonal as the building block and the
reflection of the lower triangular block is used as the environment
block. At each iteration the whole diagonal is thus added to the lower
triangular block. In this manner one of the problems with adding just
a row of sites to the system block is overcome; the lattice grows in
both spatial dimensions. Furthermore, the procedure retains a high
degree of symmetry and could, in principle, be used as an infinite
system method. The problem is, of course, that the number of states
needed to describe the exact diagonal block still increases
exponentially with the linear system size. The method was, however,
implemented.

When adding an exact diagonal to the system the energy per site
decreased as the system size was increased, until the computer ran out
of memory.  Having passed this simple test the next problem to be
addressed was the exponential increase of states in the diagonal. The
natural way to avoid the exponential growth is to select only the most
important states in the diagonal block by diagonalizing the density
matrix for the diagonal. This was done, and at each iteration a single
site was added in one corner. The local energy did, however, start to
increase as the system size was increased. As in the 1D case, the
reason seemed to be that a site was added at the boundary of the
system. If periodic boundary conditions are used this may be a
possible blocking formula, but it does not work with open boundary
conditions.

\begin{figure}
\centering
\epsfxsize=7cm
\leavevmode
\epsffile{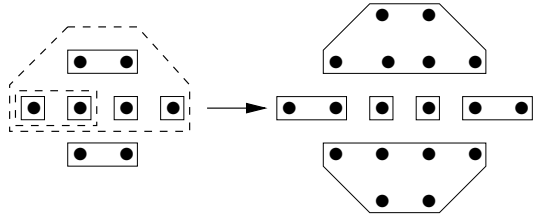}
\vskip0.5cm
\caption{A blocking algorithm for a tilted 2D lattice with an even 
number of sites. The density matrix is formed both for the diagonal 
block and the upper  triangular block (both indicated by heavy dashed 
lines).}
\label{fig:2D.2}
\end{figure}

\begin{figure}
\centering
\epsfxsize=7cm
\leavevmode
\epsffile{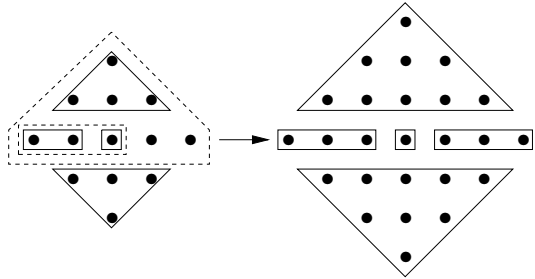}
\vskip0.5cm
\caption{A blocking algorithm for a tilted 2D lattice with an odd number 
of sites. The density matrix is formed both for the diagonal block and 
the upper triangular block (both indicated by heavy dashed lines).}
\label{fig:2D.3}
\end{figure}

A new method was therefore tried where, in analogy with the 1D method,
the diagonal was divided into two blocks with the additional site
added in the middle. The local energy did, however, still increase as
a function of lattice size. A potential problem seemed to be that when
working with square lattices, one is forced to construct the wave
function for a lattice with an even number of sites from the wave
function for a lattice with an odd number of sites. Lattices with odd
and even numbers of sites do, however, have quite different wave
functions. This issue can be avoided if one studies lattices tilted by
45 degrees.  First an attempt was made to use lattices tilted by 45
degrees containing an even number of sites, see
Fig.~\ref{fig:2D.2}. With this geometry the standard 1D DMRG technique
can be used for the diagonal, with the exception that the superblock
also contains the triangular blocks. The site energy did still not
decrease monotonically as the system size was increased. Therefore
tilted lattices with an odd number of sites were investigated, see
Fig.~\ref{fig:2D.3}, and it was found that the energy decreased
monotonically as the lattice size was increased. This was the only
blocking scheme found in this study for which this was the case. The
fact that there exists such an algorithm is certainly encouraging, and
not self-evident, as pointed out above. Since this was the most
promising algorithm found in this study the results will be analyzed
in more detail next.

\begin{figure}
\centering
\epsfxsize=9cm
\leavevmode
\epsffile{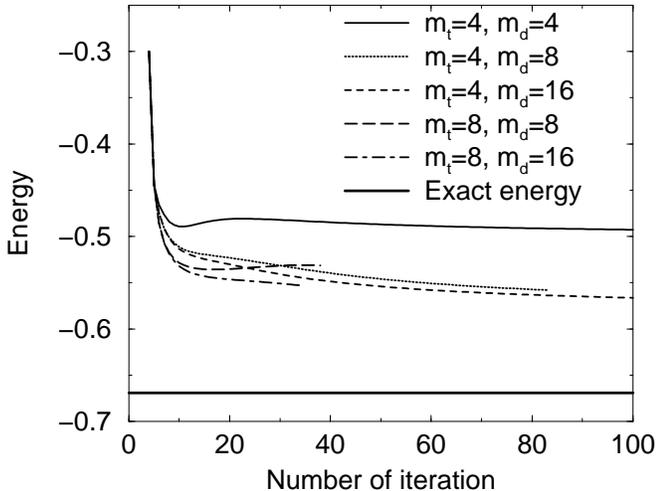}
\vskip0.5cm
\caption{The energy as a function of number of iterations for the
2D Heisenberg model. The number of states kept in the diagonal block
is denoted $m_d$, while the number of states in the triangular block is 
given by $m_t$. The exact result is obtained from Monte Carlo simulations
(Ref. 7).}
\label{fig:ene2D}
\end{figure}

In Fig.~\ref{fig:ene2D} the site energy is shown as a function of the
number of iterations. Density matrices are formed both for the
diagonal and the triangular blocks, and the number of states kept in
these blocks are denoted $m_d$ and $m_t$ respectively. The ground
state energy for the 2D Heisenberg model is -0.669437(5).\cite{Sand}
Keeping four states in the diagonal and four blocks in the triangular
block the energy levels out around -0.49. Increasing $m_d$ to 16
dramatically improves the energy to -0.57.  It seems desirable to keep
a higher number of states in the diagonal than in the triangular
block. Next the number of states in the triangular block was increased
to eight. Then only 16 states could be kept in the triangular block,
and for the number of iterations that could be done the results were
slightly better than the results obtained when keeping four states in
the triangular block and 16 states in the diagonal block.

In order for a DMRG method to be useful one has to be able to keep
enough states per block to reach convergence in the quantity studied.
Since a 2D blocking algorithm of the kind described above contains
more blocks than the traditional 1D blocking method this may prove
difficult. The superblock for the above algorithm will contain
$2m_d^2m_t^2$ states, and the density matrix for the triangular block
will contain $2m_d^2m_t$ states. The programs used in this
investigation are, however, far from optimized. Computer memory
limited the present study. By using good quantum numbers, like the
$z$-component of the spin for the Heisenberg model, all matrices
become block diagonal. In this study complete matrices were stored,
and one should be able to significantly increase the number of states
kept if only the non-zero matrix blocks are stored.

The important issue is thus to study how the bulk ground state energy
is approached as one further increases the number of states in the
blocks.  The reason for the great success of the 1D DMRG method is a
very fast convergence. If the 2D calculation shows exponential or
power law convergence one may be able to keep enough states to reach
accurate results, but if the convergence is slower this may prove
difficult.

It is also possible to make a finite system algorithm based on the
above blocking procedure. In such a method one could use the infinite
system method for the initial steps, saving both the triangular and
the diagonal blocks. Further sweeps could use these saved blocks as
environment blocks and improve the basis kept in the system blocks, in
a manner analogous to the 1D method.

The algorithm presented in this Communication bears some resemblance
to a ``four-block method'' proposed by Bursill.\cite{Burs} In both
methods the building block, which determines the growth of the system
block, does not consist of exact sites, as in the original method, but
of sites with a reduced number of states.

To conclude, using a new approach I have explored the first fully
two-dimensional infinite system DMRG calculation. The fixed point in
two-dimensions could be explicitly studied and it was shown that there
exists an algorithm for which the local energy for the Heisenberg
model decreases monotonically as the system size is increased.
Previous results indicated that it is necessary to keep a number of
states that grows exponentially with the linear system size to
maintain a certain accuracy. This does not appear to be the case with
the infinite system algorithm as the fixed point is approached. The
method preserves a high degree of the symmetry of the lattice and
could be used as a starting point for a finite system
algorithm. Further studies are necessary to verify whether the method
presented here, or other similar algorithms, exhibit convergence that
is fast enough to calculate properties of large two-dimensional
electronic systems.

I thank Steven Girvin, Claudio Gazza and Anders Sandvik for helpful
conversations. The research was supported by NSF Grants
No. CDA-9601632, DMR-9714055 and DMR-9629987. I acknowledge support
from Suomalainen Tiedeakatemia and I am thankful for the hospitality
of the University of Virginia, where part of the work was done.


\begin{references} 
\bibitem[*]{pa} Permanent address: National High Magnetic Field Laboratory,
 Tallahassee, FL 32310.
\bibitem{Dago} E. Dagotto, Rev. Mod. Phys. {\bf 66}, 763 (1994).
\bibitem{Loh} E. Y. Loh Jr., J. E. Gubernatis, R. T. Scalettar, 
              S. R. White, D. J. Scalapino and R. L. Sugar, 
              Phys. Rev. B {\bf 41}, 9301 (1990).
\bibitem{White1} S. R. White, Phys. Rev. Lett. {\bf 69}, 2863 (1992);
                              Phys. Rev. B {\bf 48}, 10345 (1993).
\bibitem{Wies} A  recent communication claims to have
              solved the sign problem in some models, see S. Chandrasekharan
              and U. Wiese, cond-mat/9902128. 
\bibitem{White3} S. R. White, Phys. Rev. Lett. {\bf 77}, 3633 (1996).
\bibitem{Jongh} M. S. L. du Croo de Jongh and J. M. J. van Leeuwen, Phys. Rev.
B {\bf 57}, 8494 (1998).
\bibitem{Sand} A. W. Sandvik, Phys. Rev. B {\bf 56}, 11678 (1997).
\bibitem{White4} S. R. White and R. M. Noack, Phys. Rev. Lett. {\bf 68}, 
                 3487 (1992).
\bibitem{White2} S. R. White and D. A. Huse, Phys. Rev. B {\bf 48},
 3844 (1993).
\bibitem{Liang} S. Liang and H. Pang, Phys. Rev. B {\bf 49}, 9214 (1994).
\bibitem{Xiang} T. Xiang, Phys. Rev. B {\bf 53}, 10445 (1996).
\bibitem{White5} S. R. White and D. J. Scalapino, Phys. Rev. Lett. {\bf 80}, 
1272 (1998).
\bibitem{Ostl} S. \"Ostlund and S. Rommer, Phys. Rev. Lett. {\bf 75},
3537 (1995).
\bibitem{Burs} R. J. Bursill, cond-mat/9812349.
\end{references}
\end{document}